\documentclass[12pt, letterpaper]{article}
\pdfoutput=1
\usepackage{amsmath}
\usepackage{amssymb}
\usepackage{graphicx}
\usepackage{color}
\usepackage{setspace}
\usepackage{float}
\usepackage{amssymb,amsfonts,amsmath}
\usepackage{setspace}
\usepackage[top=25mm, bottom=25mm, left=25mm, right=25mm]{geometry}
\usepackage{graphicx}

\usepackage{pdfpages}

\usepackage[english]{babel}

\newcommand \beqa {\begin{eqnarray} } 
\newcommand \eeqa {\end{eqnarray}}
\newcommand \beq{\begin{equation}}
\newcommand \eeq{\end{equation}}

\newcommand \eps {\epsilon}
\newcommand \la {\langle}
\newcommand \ra {\rangle}

\begin{document}

\includepdf[pages=1-22]{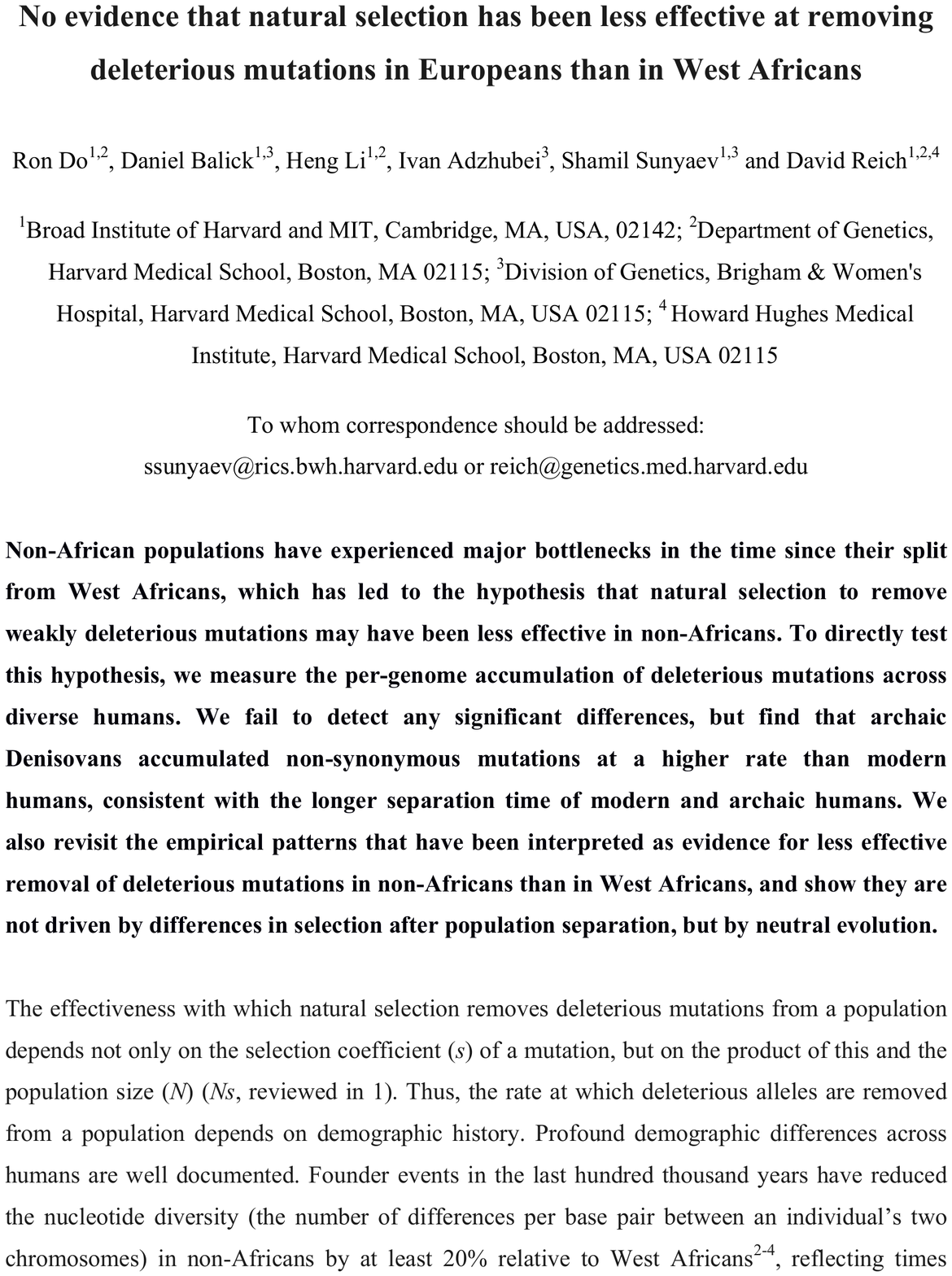}
\includepdf[pages=1-17]{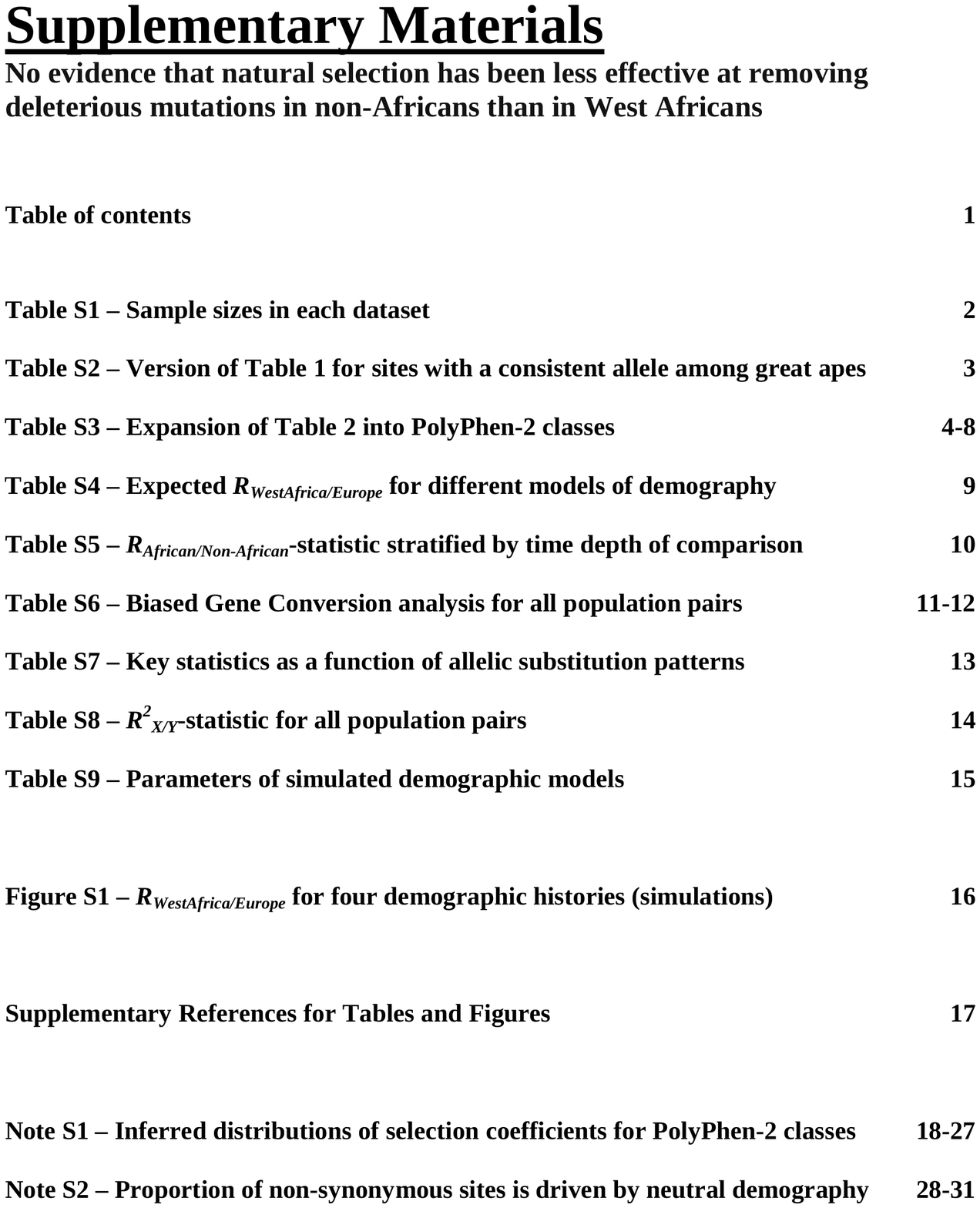}

\title{Note S1: Inferred distributions of selection coefficients for PolyPhen-2 classes}
\author{}
\date{}
\maketitle
\thispagestyle{empty}
\pagestyle{empty} 

\vspace{.8cm}

\section*{Abstract}
This note details the empirical fitting of the site frequency spectrum (SFS) from 1000 genomes data to determine the underlying distribution of fitness effects (DFE) for new mutations.  Of particular interest is the DFE for PolyPhen2 classes.


\section*{Aims and goals}

Here we describe the technique used to analyze the distribution of selective effects of de novo mutations that form the distribution of fitness effects (DFE).  Our primary aim is to infer this distribution from the site frequency spectrum of polymorphic non-synonymous alleles in the context of a given demographic history and total mutation rate.  The de novo DFE in humans is in principle independent of population history and other demographic differences between individuals, allowing us to infer the distribution from a single fixed demography without loss of generality, provided the demographic inference is accurate.

\section{Site frequency spectra}
We use coding sequences from the 1000 genomes Yoruban (YRI) and Northern Europeans from Utah (CEU) populations to create a site frequency spectrum (SFS) in the form of a minor allele frequency (MAF) spectrum for both synonymous and non-synonymous sites.  Additionally, we stratify the non-synonymous SFS by predicted PolyPhen2 classes, labeled benign, possibly damaging, and probably damaging in order of increased predicted effect.  

%
%

\subsection{Simulated MAFs}

Using the demographic inferences given in Gravel, et al.\cite{Gravel2011}, we simulate a genome of length 100Mb through the inferred demographic histories of European and African populations for a range of selective effects.  In particular, the simulator tracks the derived allele frequencies of $10^8$ independently evolving sites, in the infinite recombination limit with no linkage.  Mutations are introduced at a rate $\mu~=~2~\times~10^{-8}$ per site per individual per generation.  The population size is time dependent and reflects the demography associated with the population of interest.  After completing roughly $5000$ generations of recent demographic history, the allele frequencies are subsampled to the sample size of the associated 1000 genomes population sample, 88 for YRI and 85 for CEU.  The results of this simulation provide expectations for the MAF for alleles with a single selective coefficient $s$.  We simulate separately for $s = \{0, -10^{-3}, -10^{-2}\}$, which we consider to be neutral, weakly deleterious, and strongly deleterious, respectively.  These  selective coefficients are chosen to represent the range of realistic selective effects expected to be segregating in the human population.  Alleles of stronger selective effect are likely to be absent in all but the largest population samples, and will be incorporated into the $s=-10^{-2}$ fitness class in our fit.  
These simulated MAFs provide the basis for our fit, as we will estimate the coefficients of their linear combination to determine the DFE.

\section{Overall scale and target size}

The number of bases simulated clearly overestimates the length of the human coding genome.  The total coding genome is thought to be roughly $30Mb$ long, accounting for about $1\%$ of the whole genome.  Since estimates of both the mutation rate and target size are known to be relatively imprecise, we use the synonymous MAF to determine the overall rescaling for fitting our simulations to 1000 genomes data.  Additionally, this method accounts for coverage issues, etc., assuming the same fraction of synonymous and non-synonymous sites are affected.

\subsection{Scale factor for synonymous sites}

Assuming synonymous sites are selectively neutral, we use a maximum likelihood fit with a single parameter to determine the scale factor for synonymous sites.  The log likelihood is calculated as follows.
\beq
\log \mathcal L = \sum_{i=1}^{N} (D_i \log [F_i ] - F_i)
\eeq
Here $D_i$ represents the $i^{th}$ bin of the MAF from data, where $i \in [1,N]$ corresponds to allele count in the sample ranging from singletons at frequency $x = i/2N=1/2N$ to alleles present in half of the haploid individuals at $x = N/2N = 1/2$.  Similarly, $F_i$ corresponds to counts in the fit to simulation, and is a function of fit parameters $\eps_k$.  For the present purposes, we are interested in determining the maximum likelihood for the following form of $F_i(\eps)$.
\beq
F_i(\eps) = \eps \ S^{i}_0
\eeq
$S^0_i$ represents the $i^{th}$ count of the MAF for the appropriately down-sampled neutral simulation with $s = 0$.  The maximum log likelihood is given by the following expression.
\beq
\max[\log \mathcal L(\eps^{syn})] = \max\left[  \sum_{i=1}^{N} (D_i \log [\eps^{syn} S^i_0 ] - \eps^{syn} S^i_0)  \right]
\eeq
We use the YRI synonymous MAF $D^{YRIsyn}_i$ and the simulated YRI MAF for $s=0$ to determine $\eps^{YRIsyn}$ numerically.  
The synonymous scale factor for YRI is determined by the maximum log likelihood value at $\eps^{YRIsyn} = 0.093$.  Analogously, the synonymous scale factor for CEU has a maximum log likelihood value of $\eps^{CEUsyn} = 0.097$.

\subsection{Scale factor for non-synonymous sites}

Kryokov, et al. \cite{Kryukov2009} estimates the synonymous and non-synonymous fractions of the coding genome to be $0.32$ and $0.68$, respectively.  This can be used to determined the appropriate scale factor for non-synonymous sites.  The scale factor is simply the ratio of the total mutation rate in the target to the total simulated mutation rate.
\beq
\epsilon^{syn} = \frac{U_{syn}^{data}}{U^{sim}} =  \frac{(\mu L_{syn})}{U^{sim}}
\eeq
This can be solved for $\mu$ and substituted in to the non-synonymous expression to determine the non-synonymous scale factor.
\beqa
\eps^{nonsyn} &=& \frac{U_{nonsyn}^{data}}{U^{sim}} =  \frac{(\mu L_{nonsyn})}{U^{sim}} \nonumber \\
\nonumber \\
&=&  \frac{ L_{nonsyn}}{L_{syn}} \eps^{syn} =  \left(\frac{68}{32}\right) \eps^{syn}
\eeqa
We find the following scale factors for the YRI and CEU simulated data.  
\begin{center}
    \begin{tabular}{ | l | l | l | p{5cm} |}
    \hline
	$\eps^{YRI}_{nonsyn}  $  &    $  \eps^{CEU}_{nonsyn} $ \\ \hline
	0.198 & 0.207   \\ \hline
    \end{tabular}
\end{center}

\subsection{Scale factors for Polyphen2 classes}

The Polyphen2 software provides functional predictions that can be stratified into 3 classes: benign, possibly damaging, and probably damaging.  One can compute the target size of these classes as a fraction of the total non-synonymous coding genome.  This is accomplished by enumerating all possible point mutations from the hg19 human reference genome and classifying each mutation.  We use the context dependent $64\times 4$ weight matrix of single point mutations from a given triplet to all others \cite{Asthana2007}.  Each of the $4^3$ possible triplets has an associated matrix.  Using HumVar, we compute approximate fractions for PolyPhen2 classes found in the following table.  
\begin{center}
    \begin{tabular}{ | l | l | l | p{5cm} |}  
    \hline
    prediction &        fraction (\%) \\ \hline
     benign        &  50.0   \\ \hline
     possibly damaging \  & 16.7 \\ \hline
    probably damaging     & 33.3 \\ \hline
    unknown*     &    $\ll$ 1 \\ \hline
    \end{tabular}
\end{center}
To confirm that this estimate is not biased by ancestry or recent demography, we stratify the human reference genome by predicted ancestry and find no substantial difference from these approximate values.  From these fractions, we compute the appropriate scale factors for our fitting procedure.
\begin{center}
    \begin{tabular}{ | l | l | l | p{5cm} |}
    \hline
 $ \eps_{benign}^{YRI}$  &   $\eps_{possibly}^{YRI} $ &  $  \eps_{probably}^{YRI}$ \\ \hline
 0.099 &   0.033  &  0.066   \\ \hline
    \end{tabular}
\end{center}
\begin{center}
    \begin{tabular}{ | l | l | l | p{5cm} |}
    \hline
$\eps_{benign}^{CEU} $  &   $\eps_{possibly}^{CEU} $ &  $  \eps_{probably}^{CEU}$ \\ \hline
 0.104 &   0.035  &  0.069   \\ \hline
    \end{tabular}
\end{center}

\section{Maximum Likelihood fit}

Using the scale factors determined in the previous section, we compute the maximum log likelihood for a linear combination of selective effects.  For simplicity, we choose to represent the DFE as a sum of several single $s$ effect classes, rather than using a continuous functional form.  We acknowledge that this three point mass model is a simplification of the true distribution of selection coefficients, but believe that it is useful for the purpose of obtaining a rough prediction of the expected value of the R-statistic for specific PolyPhen-2 classes.
\beq
\log \mathcal L(\{\alpha_k\}) = \sum_{i=1}^{N} \left(D_i \log [F_i (\{\alpha_k\})] - F_i(\{\alpha_k\}) \right)
\eeq
  We use the following form for the fit function $F(\{\alpha_k\})$.
\beq
F_i(\{\alpha_k\}) = \eps^{nonsyn} \sum_k \alpha_k  \ S_{k}^i = \eps^{nonsyn} \left(  \alpha_0  \ S_{0}^i  + \alpha_3  \ S_{3}^i +\alpha_2  \ S_{2}^i  \right)
\eeq
We employ the notation $k = 0$ for the simulated $s=0$ MAF, $k = 3$ for the simulated $s=-10^{-3}$ MAF, and  $k = 2$ for the simulated $s=-10^{-2}$ MAF.  In this form, $S_i^3$ represents the MAF for the weakly selected sites, and $\alpha_3$ is the fraction of the DFE that falls into this category.  By estimating the maximum likelihood we can re-assemble the DFE in a rudimentary form as a fraction of mutations that fall into the category of neutral, weakly deleterious, and strongly deleterious.  Since the overall scale factor is fixed, the $\alpha_k$ coefficients must be normalized with the following constraint.
\beq
\sum_k \alpha_k = 1
\eeq
This restricts the fit function as follows.
\beq
F_i(\{\alpha_k\}) = \eps^{nonsyn} \left(  \alpha_0  \ S_{0}^i + \alpha_3  \ S_{3}^i +(1-\alpha_0 - \alpha_3) \ S_{2}^i  \right)
\eeq
Note that for the present purposes, we have chosen 2 free parameters to fit, such that $\{\alpha_k\}=~\{\alpha_0,\alpha_3\}$.  For a 3 parameter fit with an additional nearly neutral class at $s=-10^{-4}$, for example, we simply introduce $\alpha_4$ and $S_4^i$ and modify the constraint $(\alpha_0 + \alpha_3 + \alpha_3 + \alpha_2) =1$, with free parameters $\{\alpha_k\}=\{\alpha_0,\alpha_4,\alpha_3\}$.  This method can be easily extended to fit an arbitrary number of parameters by including additional $S_k^i$ for various selective effects.  We have found this unnecessary for the present purposes, as it results in the effective overfitting of the DFE.

The maximum likelihood fit for 2 parameters is given simply by the following equations.
\beqa
\max \left[ \log \mathcal L(\alpha_0, \alpha_3) \right] =  \max \left[ \sum_{i=1}^{N} \left(D_i \log [F_i (\{\alpha_k\})] - F_i(\{\alpha_k\}) \right)  \right] \\ \nonumber \\
F_i(\{\alpha_k\}) = \eps^{nonsyn} \left(  \alpha_0  \ S_{0}^i + \alpha_3  \ S_{3}^i +(1-\alpha_0  - \alpha_3) \ S_{2}^i  \right)
\eeqa

\section{Results}
\label{results}

Using the method outlined above, we compute the maximum likelihood fits for various PolyPhen2 classes using YRI, CEU, and a joint measure that is the sum of the log likelihood functions of both YRI and CEU.  Since the DFE should in principle be independent of demographic history, one can use the overlap of the independent measures in YRI and CEU in the form of the joint log likelihood (defined as a sum of the two log likelihoods) to produce a fit that is less sensitive to demographic errors in either of the two populations individually.  The maximum log likelihood fit is summarized in the tables below.  Errors are given for the joint fit, as this will be used in our subsequent analysis. 
\begin{table}[H]
    \begin{tabular}{|l|l|l|l|}
    \hline
    {\bf 2 parameter fit (YRI)} & neutral ($s=0$) & weak ($s=-10^{-3}$) & strong ($s=-10^{-2}$) \\ \hline
    all non-synonymous         & 0.20          & 0.44                           & 0.36                             \\ \hline
    benign                    & 0.28          & 0.56                           & 0.16                             \\ \hline
    possibly damaging         & 0.17          & 0.50                           & 0.34                             \\ \hline
    probably damaging         & 0.09          & 0.25                           & 0.66                             \\ \hline
    \end{tabular}
\end{table}
\begin{table}[H]
    \begin{tabular}{|l|l|l|l|}
    \hline
    {\bf 2 parameter fit (CEU)} & neutral ($s=0$) & weak ($s=-10^{-3}$) & strong ($s=-10^{-2}$) \\ \hline
    all non-synonymous         & 0.18          & 0.55                           & 0.27                             \\ \hline
    benign                    & 0.26          & 0.68                           & 0.06                             \\ \hline
  possibly damaging         & 0.15          & 0.63                           & 0.21                             \\ \hline
   probably damaging         & 0.08          & 0.32                           & 0.60                             \\ \hline
    \end{tabular}
\end{table}
\begin{table}[H]
    \begin{tabular}{|l|l|l|l|}
    \hline
    {\bf 2 parameter fit (Joint)} & neutral ($s=0$) & weak ($s=-10^{-3}$) & strong ($s=-10^{-2}$) \\ \hline
    all non-synonymous         & 0.19  $\pm$ 0.01        & 0.47     $\pm$ 0.04                      & 0.33   $\pm$ 0.05                          \\ \hline
    benign                    & 0.27       $\pm$ 0.02   & 0.60     $\pm$ 0.07                      & 0.13       $\pm$ 0.07                      \\ \hline
  possibly damaging         & 0.16  $\pm$ 0.03        & 0.54    $\pm$ 0.11                      & 0.29    $\pm$ 0.11                         \\ \hline
   probably damaging         & 0.09    $\pm$ 0.01      & 0.27   $\pm$ 0.06                        & 0.64    $\pm$ 0.06                         \\ \hline
    \end{tabular}
\end{table}

\subsection{Log Likelihood plots}

\begin{figure}[H]
\begin{center}
\includegraphics[width=0.5\columnwidth]{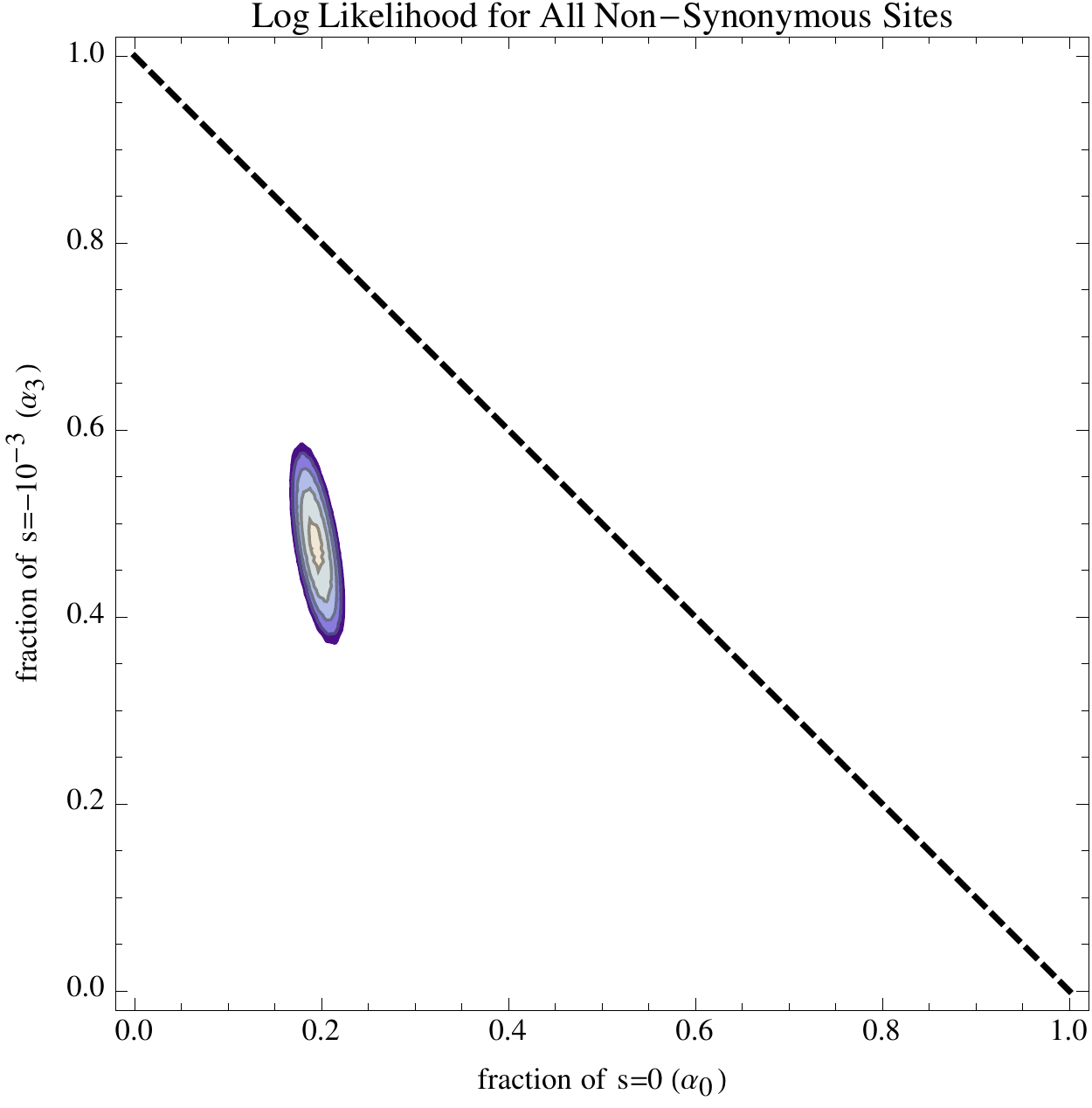}
\end{center}
\caption{\label{ML2nonsyn}  The log likelihood plot for the joint inference from YRI and CEU data for all non-synonymous sites is shown for a two parameter fit.  Contours are plotted representing two standard deviations from the peak.  The coefficients of $s=0$ and $s=-10^{-3}$, represented as ($\alpha_0$,$\alpha_3$), are plotted on the $x$ and $y$ axes, respectively.  The fraction of strongly deleterious ($s=-10^{-2}$) sites in the DFE is constrained by the equation $\alpha_0 + \alpha_3 + \alpha_2 = 1$.  This constraint restricts allowed values to below the dashed line. The maximum likelihood fit is located at $\{\alpha_0, \alpha_3, \alpha_2\} = \{0.19, 0.47, 0.33\}$.}
\end{figure}

\begin{figure}[H]
\begin{center}
\includegraphics[width=0.32\columnwidth]{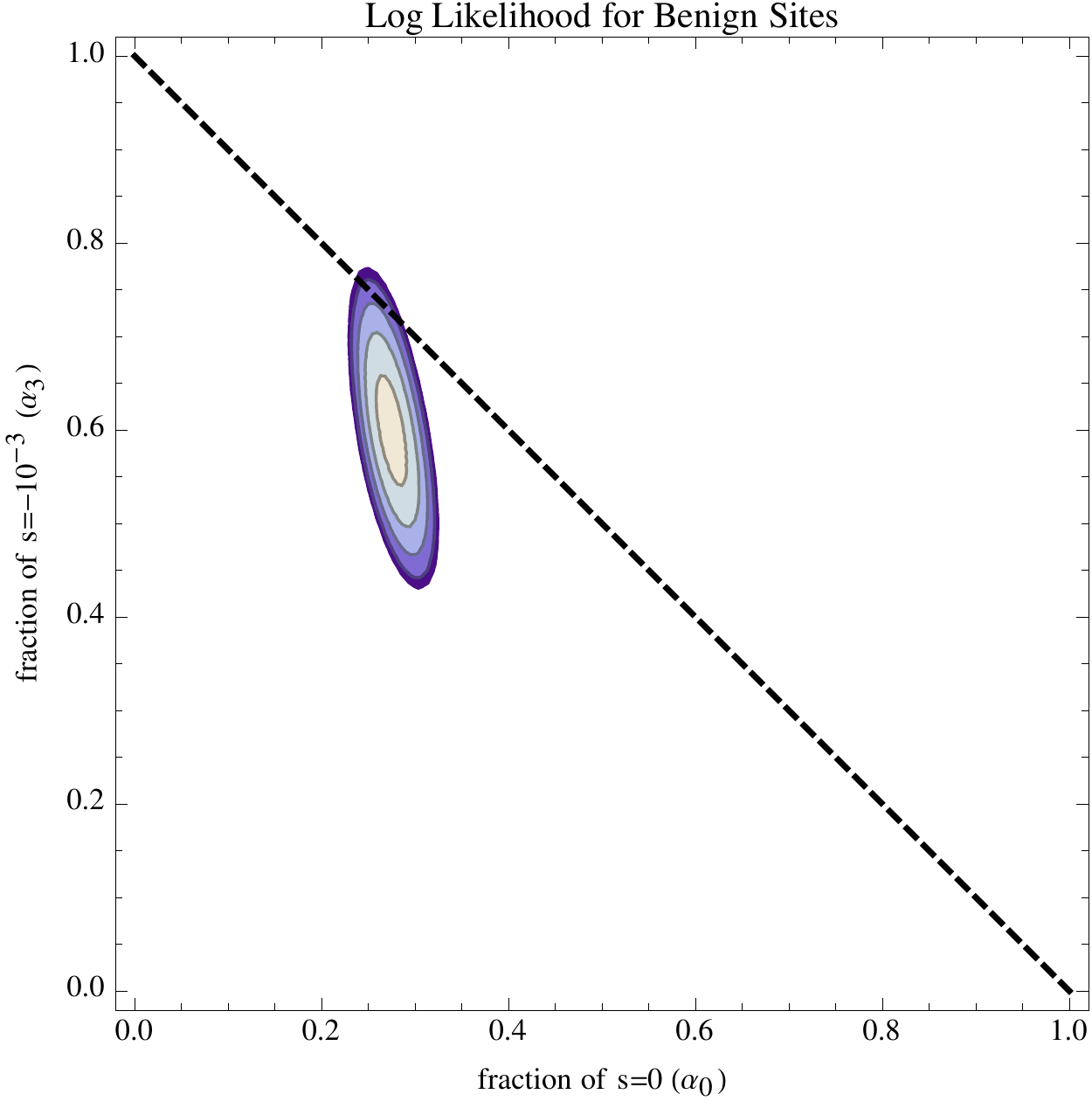}
\includegraphics[width=0.32\columnwidth]{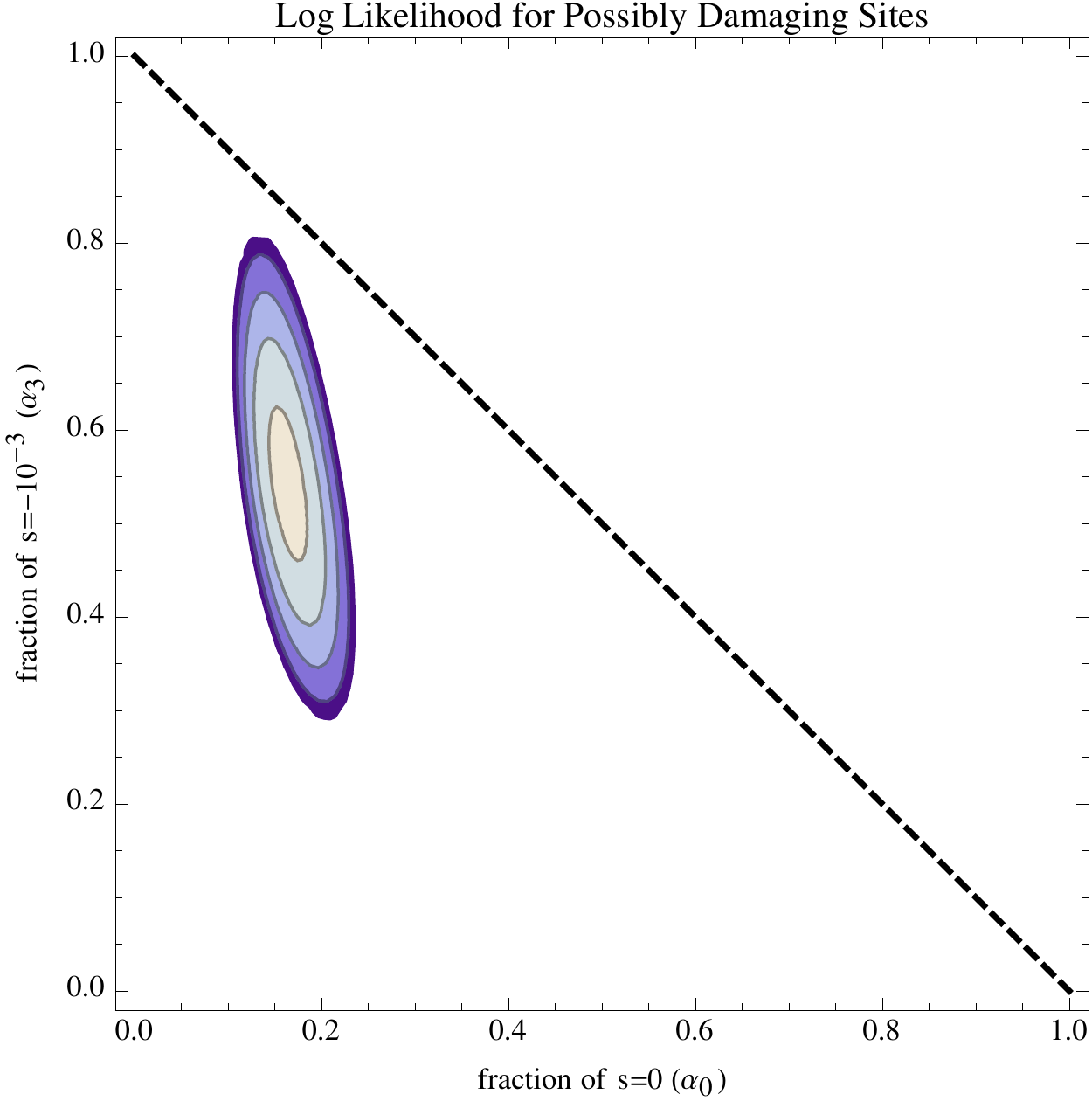}
\includegraphics[width=0.32\columnwidth]{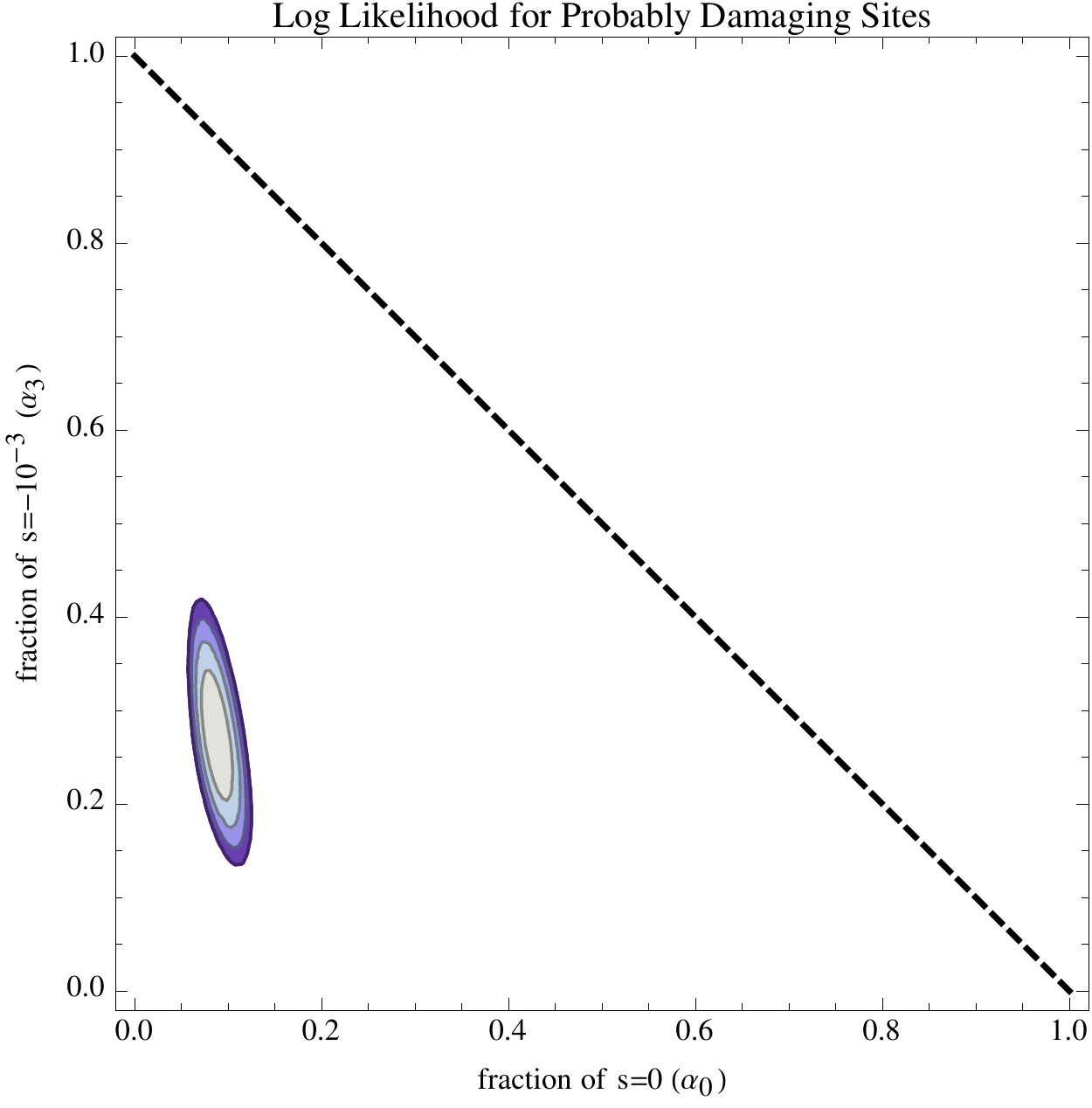}
\end{center}
\caption{\label{ML2poly} Log Likelihood plots for the 2 parameter fit from the joint inference of YRI and CEU data are plotted for PolyPhen2 classes.   { \bf LEFT:} Benign sites.   { \bf MIDDLE:} Possibly damaging sites. { \bf RIGHT:} Probably damaging sites.   All plots have axes ($\alpha_0$,$\alpha_3$) corresponding to neutral and weakly deleterious alleles and display two standard deviations from the maximum.  The constraint $\alpha_0 + \alpha_3+ \alpha_2 = 1$ is satisfied, and only values below the dashed line are allowed. Note that the fit favors smaller fractions of neutral and weakly deleterious sites in favor of strongly deleterious sites with increasing PolyPhen2 score, consistent with prediction.}
\end{figure}

The log likelihood surface for the two parameter fit can be visualized in a contour plot shown in Figure \ref{ML2nonsyn}.   
We note that the normalization condition $\sum_k \alpha_k = 1$ determines the strongly deleterious class uniquely.  Figure  \ref{ML2poly} plots log likelihood contours for the benign, possibly damaging, and probably damaging PolyPhen2 classes.  We note a trend in the location of the maximum towards smaller values with increased predicted effect.  All of the mass that vanishes in this process contributes to enhancing the weight of the strongly deleterious class.  This is consistent with the stratification by PolyPhen2 score, reinforcing our results.  

\section{Using the DFE to appropriately weight $R$}

Here we use the inferred distribution of fitness effects, $\rho(s)$, to define an expected value $\langle R \rangle$ corresponding to the value of $R$ that we expect to observe in population data.  The appropriately weighted mutation load $\la L \ra$ for a given population is given by convoluting the load at different $s$ values over the DFE.
\beq
\la L \ra  = \int ds \ \rho(s)  \ L(s)
\eeq
This is true for both populations independently, since the DFE is roughly the same, allowing us to compute the expected $\la R \ra$ as follows.
\beq
\la R \ra = \frac{ \la L \ra_{pop0}  }{\la L \ra_{pop1}  } = \frac{ \int ds \ \rho(s)  \  L^{pop0}(s) }{ \int ds \ \rho(s)   \  L^{pop1}(s) }   
\eeq
For the discretization of the DFE into neutral, weakly deleterious, and strongly deleterious components, this can be rewritten as the following sum.
\beqa
\la R \ra & =& \frac{ \sum_k \alpha_k L ^{pop0}(s_k) }{  \sum_k \alpha_k L ^{pop1}(s_k)  }   \nonumber \\ \nonumber \\
&= &\frac{ \alpha_0 L ^{pop0}(s=0)+\alpha_3 L ^{pop0}(s=-10^{-3})+\alpha_2 L ^{pop0}(s=-10^{-2}) }{  \alpha_0 L ^{pop1}(s=0)+\alpha_3 L ^{pop1}(s=-10^{-3})+\alpha_2 L ^{pop1}(s=-10^{-2})  }   
\eeqa
Here the ${\alpha_k}$ correspond to the fractions given in the results table above, and can represent appropriate values for all non-synonymous sites, or those for any of the PolyPhen2 classes.

\subsection{Computing  $\la R \ra$, the weighted $R$ statistic}

Here, we calculate a weighted mutation load for population 0 (African) and population 1 (European) using fractions obtained from the maximum likelihood fits from the inferred distribution of fitness effects from Section \ref{results} and from simulated mutation loads for average selection coefficients $s=\{0, -0.001, -0.01\}$.  We calculated the weighted $R$ statistic, denoted $\la R \ra$,  as the ratio of the weighted mutation loads corresponding to population 0 and population 1.  We calculate $ \la R \ra$ for all non-synonymous sites, in addition to Polyphen classes, including benign, possibly damaging, and probably damaging sites (see tables below).

We calculated the expected $\la R \ra$  from simulations for four demographic models: Tennessen \cite{Tennessen2012}, Gravel \cite{Gravel2011}, Lohmueller \cite{Lohmueller2008}, and a simple bottleneck without exponential growth.  We compare $\la R \ra$ from simulations with the $R$ statistic observed in African Americans/European Americans from the Exome Sequencing Project (ESP) to assess the validity of different demographic models.  Using this approach, we are unable to reject the Tennessen, Gravel and Lohmueller models, since $\la R \ra$  from these models are all within the 95\% confidence intervals of $R$ from ESP for all classes.  The square bottleneck prediction is $2.09$ standard errors from the empirical observation from the ESP measurement which is weakly suggestive that this model is not consistent with the data.  These results suggest that this approach, the accumulation of deleterious mutations in two populations, along with the inferred DFE, can be a useful tool to evaluate the validity of different demographic models.
\begin{center}
\begin{table}[H]
    \begin{tabular}{|l|l|l|l|}
    \hline
    {\bf all non-synonymous sites } &  $\la L \ra_{pop0} $ & $\la L \ra_{pop1} $ & $\la R \ra$  \\ \hline
    Tennessen         & 0.000139         &  0.000140                           &  0.989                             \\ \hline
    Gravel                  &  0.000138          &  0.000140                           &  0.987                             \\ \hline
  Lohmueller         &  0.000113          &  0.000114				&  0.992                             \\ \hline
   Simple Bottleneck         & 0.000138          & 0.000141                           & 0.978                             \\ \hline
    \end{tabular}
\end{table}
\end{center}

\begin{table}[H]
    \begin{tabular}{|l|l|l|l|}
    \hline
    {\bf benign } &  $\la L \ra_{pop0} $ & $\la L \ra_{pop1} $ & $\la R \ra$  \\ \hline
    Tennessen         & 0.000193         &  0.000195                   	        &  0.990                             \\ \hline
    Gravel                  &  0.000192          &  0.000194         	         	   &  0.988                           \\ \hline
  Lohmueller         &  0.000157			&  0.000158			&  0.993					\\ \hline
   Simple Bottleneck         & 0.000192          & 0.000196                           & 0.979                             \\ \hline
    \end{tabular}
\end{table}

\begin{table}[H]
    \begin{tabular}{|l|l|l|l|}
    \hline
    {\bf possibly damaging } &  $\la L \ra_{pop0} $ & $\la L \ra_{pop1} $ & $\la R \ra$  \\ \hline
    Tennessen         & 0.000123         &  0.000125                       & 0.985                             \\ \hline
    Gravel                  &  0.000123          &  0.000125                           &  0.984                          \\ \hline
  Lohmueller         &  0.000101          &  0.000103                         &  0.989                             \\ \hline
   Simple Bottleneck         & 0.000123       & 0.000126                          & 0.973                            \\ \hline
    \end{tabular}
\end{table}

\begin{table}[H]
    \begin{tabular}{|l|l|l|l|}
    \hline
    {\bf probably damaging } &  $\la L \ra_{pop0} $ & $\la L \ra_{pop1} $ & $\la R \ra$  \\ 	\hline
    Tennessen & 0.00006909          &  0.00006995                            &  0.988                             \\ \hline
    Gravel & 0.00006887 & 0.00006982 & 0.986 \\ \hline
    Lohmueller & 0.00005698 & 0.00005748 & 0.991 \\ \hline
        Simple Bottleneck & 0.00006885 & 0.00007048 & 0.977 \\ \hline
    \end{tabular}
\end{table}

\newpage
\singlespacing
\bibliography{_appendix_on_DFE}

\includepdf[pages=1-4]{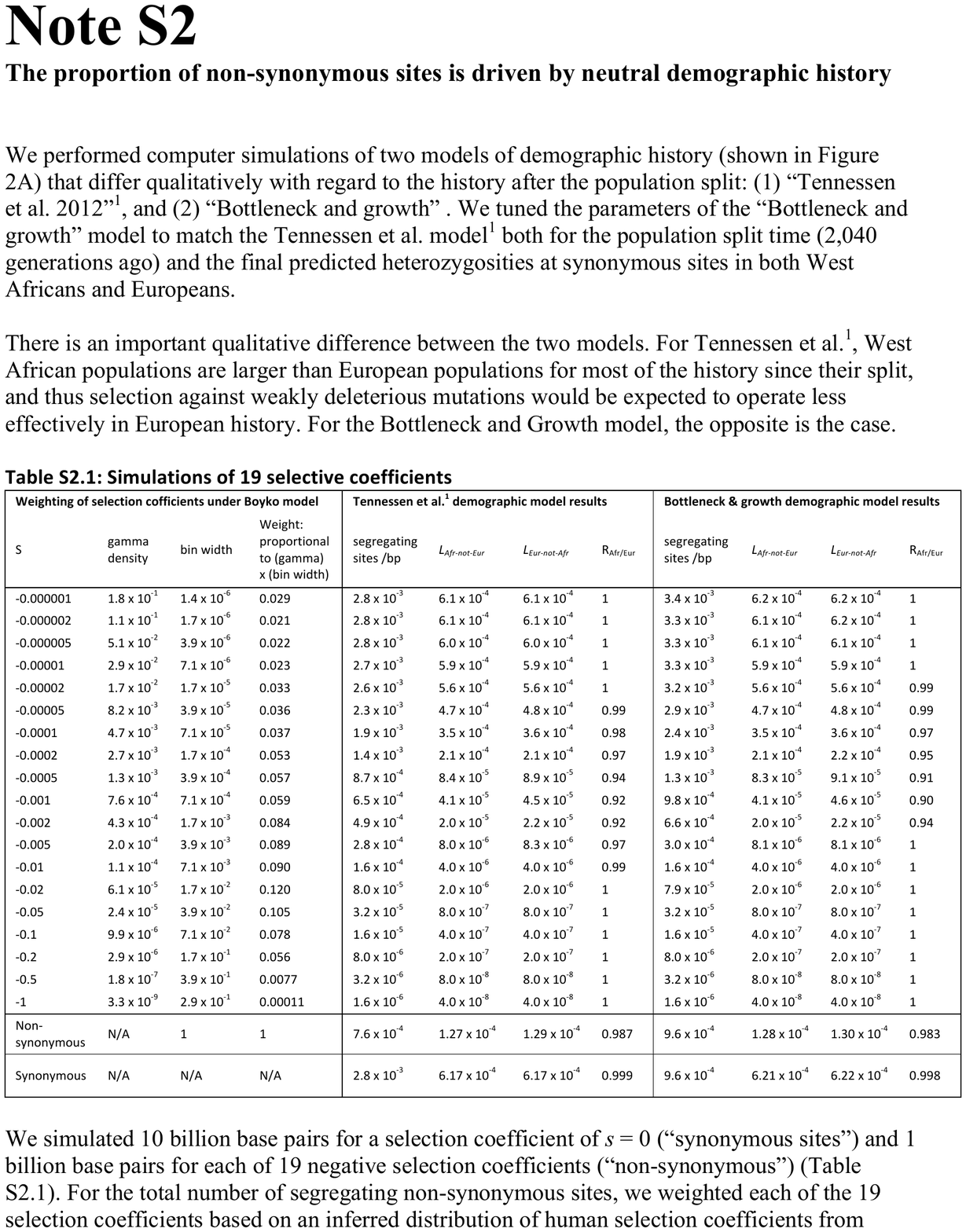}

\end{document}